\documentclass[lettersize,journal]{IEEEtran}

\usepackage{amsmath,amsfonts,amssymb}
\usepackage{algorithmic}
\usepackage{algorithm}
\usepackage{array}
\usepackage[caption=false,font=normalsize,labelfont=sf,textfont=sf]{subfig}
\usepackage{textcomp}
\usepackage{stfloats}
\usepackage{url}
\usepackage{verbatim}
\usepackage{graphicx}
\usepackage{cite}
\usepackage{cite}
\usepackage{url}								% Proper URL formatting
\usepackage{siunitx}							% Proper SI unit formatting
\usepackage{booktabs}							% Varying horizontal rule thickness, improved spacing
\usepackage{xltabular}
\usepackage{tabularx}
\newcolumntype{Y}{>{\centering\arraybackslash}X} % Centering column that uses full width
\newcolumntype{Z}{>{\raggedleft\arraybackslash}X} % Centering column that uses full width
\usepackage{enumitem} 
\usepackage{makecell}							% Table cells for holding content
\usepackage{multirow}							% Multi-row cells in tables
\usepackage[display-foreign=false]{acro}			% Acronyms (arg. for plural formatting fix)
\usepackage{gensymb}
\usepackage{upgreek}							% Used for upright greek characters (e.g., in units)
\acsetup{single}								% Do not define acro's only used once
\usepackage{mathtools}
\usepackage{bm}
\usepackage{yhmath}
\usepackage{physics}
\usepackage{xfrac}
\usepackage{color}
\usepackage{xcolor}
\DeclareAcronym{ota}{
	short = OTA,
	long = over-the-air
}
\DeclareAcronym{rrc}{
	short = RRC,
	long = root-raised-cosine
}
\DeclareAcronym{evm}{
	short = EVM,
	long = error vector magnitude
}
\DeclareAcronym{ser}{
	short = SER,
	long = symbol error ratio
}

\DeclareAcronym{psk}{
	short = PSK,
	long = phase-shift keying
}
\DeclareAcronym{qpsk}{
	short = QPSK,
	long = quadrature phase-shift keying
}
\DeclareAcronym{bpsk}{
	short = BPSK,
	long = binary phase-shift keying
}
\DeclareAcronym{zc}{
	short = ZC,
	long = Zadoff--Chu
}

\DeclareAcronym{ni}{
	short = NI,
	long = National Instruments
}
\DeclareAcronym{stft}{
	short = STFT,
	long = short-time Fourier Transform
}
\DeclareAcronym{bt}{
	short = BT,
	long = binary threshold
}
\DeclareAcronym{mbp}{
	short = MBP,
	long = multi-baseline time-frequency product
}
\DeclareAcronym{aoa}{
	short = AoA,
	long = angle of arrival,
}
\DeclareAcronym{si}{
	short = SI,
	long = International System of Units,
}
\DeclareAcronym{pec}{
	short = PEC,
	long = perfect electrical conductor,
}
\DeclareAcronym{iot}{
	short = IoT,
	long = internet of things
}
\DeclareAcronym{pmc}{
	short = PMC,
	long = perfect magnetic conductor
}
\DeclareAcronym{tem}{
	short = TEM,
	long = transverse electromagnetic
}

\DeclareAcronym{awgn}{
	short = AWGN,
	long = additive white Gaussian noise,
}
\DeclareAcronym{dar}{
	short = DAR,
	long = distributed aperture radar,
}
\DeclareAcronym{csi}{
	short = CSI,
	long = channel state information,
}
\DeclareAcronym{rfe}{
	short = RFFE,
	long = radio frequency front-end,
}
\DeclareAcronym{cwtt}{
	short = CWTT,
	long = continuous-wave two-tone,
}
\DeclareAcronym{ipc}{
	short = IPC,
	long = interprocess communication,
}
\DeclareAcronym{awg}{
	short = AWG,
	long = arbitrary waveform generator,
}
\DeclareAcronym{tbp}{
	short = TBP,
	long = time--bandwidth product,
}

\DeclareAcronym{toa}{
	short = ToA,
	long = time of arrival,
	long-plural = times of arrival,
}
\DeclareAcronym{gpu}{
	short = GPU,
	long = graphics processing unit,
}
\DeclareAcronym{pttsf}{
	short = PTTSF,
	long = pulsed two-tone stepped frequency,
}
\DeclareAcronym{ptt}{
	short = PTT,
	long = pulsed two-tone,
}
\DeclareAcronym{tdma}{
	short = TDMA,
	long = time division multiple access,
}
\DeclareAcronym{fdma}{
	short = FDMA,
	long = frequency division multiple access,
}

\DeclareAcronym{ue}{
	short = UE,
	long = user equipment,
}
\DeclareAcronym{bs}{
	short = BS,
	long = base station,
}

\DeclareAcronym{psll}{
	short = PSLL,
	long = peak-to-sidelobe level,
}
\DeclareAcronym{ugv}{
	short = UGV,
	long = unmanned ground vehicle,
}
\DeclareAcronym{uav}{
	short = UAV,
	long = unmanned aerial vehicle,
}

\DeclareAcronym{adas}{
	short = ADAS,
	long = advanced driver-assistance systems,
}

\DeclareAcronym{utc}{
	short = UTC,
	long = Coordinated Universal Time,
}
\DeclareAcronym{oot}{
	short = OOT,
	long = out-of-tree,
}
\DeclareAcronym{nlos}{
	short = NLoS,
	long = non-line-of-sight,
}
\DeclareAcronym{imd}{
	short = IMD,
	long = intermodulation distortion,
}
\DeclareAcronym{fm}{
	short = FM,
	long = frequency modulation,
}
\DeclareAcronym{pm}{
	short = PM,
	long = phase modulation,
}

\DeclareAcronym{xo}{
	short = XO,
	long = crystal oscillator,
}
\DeclareAcronym{tcxo}{
	short = TCXO,
	long = temperature compensated \acl{xo},
}
\DeclareAcronym{ocxo}{
	short = OCXO,
	long = oven controlled \acl{xo},
}

\DeclareAcronym{dsping}{
	short = DSP,
	long = digital signal processing
}
\DeclareAcronym{dsp}{
	short = DSP,
	long = digital signal processor
}

\DeclareAcronym{cfo}{
	short = CFO,
	long = carrier frequency offset
}
\DeclareAcronym{sfo}{
	short = SFO,
	long = sampling frequency offset
}
 \DeclareAcronym{ls}{
 	short = LS,
 	long = least-squares
 }
\DeclareAcronym{qls}{
	short = QLS,
	long = quadratic least-squares
}
 \DeclareAcronym{sinc-ls}{
 	short = sinc-LS,
 	long = sinc nonlinear least-squares
 }
 \DeclareAcronym{mf-ls}{
 	short = MFLS,
 	long = matched filter least-squares
 }
\DeclareAcronym{rf}{
	short = RF,
	long = radio frequency
}
\DeclareAcronym{lfm}{
	short = LFM,
	long = linear frequency modulation
}
 \DeclareAcronym{prf}{
 	short = PRF,
 	long = pulse repetition frequency
 }
 \DeclareAcronym{pri}{
 	short = PRI,
 	long = pulse repetition interval
 }
 \DeclareAcronym{fmcw}{
 	short = FMCW,
 	long = frequency modulated continuous-wave
 }
 \DeclareAcronym{lfmcw}{
 	short = LFMCW,
 	long = linear frequency modulated continuous-wave
 }
\DeclareAcronym{cw}{
	short = CW,
	long = continuous-wave
}
 \DeclareAcronym{dbf}{
 	short = DBF,
 	long = digital beamforming
 }
 \DeclareAcronym{sar}{
 	short = SAR,
 	long = synthetic aperture radar
 }
 \DeclareAcronym{psr}{
 	short = PSR,
 	long = point scatterer response
 }
 \DeclareAcronym{rcs}{
 	short = RCS,
 	long = radar cross-section
 }
\DeclareAcronym{crlb}{
	short = CRLB,
	long = Cramér-Rao lower bound
}
 \DeclareAcronym{dof}{
 	short = DoF,
 	long = degree of freedom
 }
\DeclareAcronym{snr}{
	short = SNR,
	long = signal-to-noise ratio
}
 \DeclareAcronym{sinr}{
 	short = SINR,
 	long = signal-to-interference-plus-noise ratio
 }
 \DeclareAcronym{fft}{
 	short = FFT,
 	long = fast Fourier transform,
 }
 \DeclareAcronym{ifft}{
 	short = IFFT,
 	long = inverse \ac{fft},
 }
 \DeclareAcronym{ift}{
 	short = IFT,
 	long = inverse Fourier transform,
 }
 \DeclareAcronym{ft}{
 	short = FT,
 	long = Fourier transform,
 }
 \DeclareAcronym{rms}{
 	short = RMS,
 	long = root-mean-square
 }
 \DeclareAcronym{rmse}{
 	short = RMSE,
 	long = root-mean-square error
 }
 \DeclareAcronym{psd}{
 	short = PSD,
 	long = power spectral density
 }
 \DeclareAcronym{rca}{
 	short = RCA,
 	long = range of closest approach
 }
 \DeclareAcronym{rda}{
 	short = RDA,
 	long = Range-Doppler Algorithm
 }
 \DeclareAcronym{rma}{
 	short = RMA,
 	long = Range Migration Algorithm
 }
 \DeclareAcronym{pfa}{
 	short = PFA,
 	long = Polar Formatting Algorithm
 }
 \DeclareAcronym{bpa}{
 	short = BPA,
 	long = Backprojection Algorithm
 }
 \DeclareAcronym{rvp}{
 	short = RVP,
 	long = residual video phase
 }
 \DeclareAcronym{jrc}{
 	short = JRC,
 	long = joint radar-communications
 }
 \DeclareAcronym{doa}{
 	short = DOA,
 	long = direction of arrival
 }
 \DeclareAcronym{hci}{
 	short = HCI,
 	long = human-computer interaction
 }
 \DeclareAcronym{its}{
 	short = ITS,
 	long = intelligent transportation systems
 }
 \DeclareAcronym{rtk}{
 	short = RTK,
 	long = real-time kinematic
 }
 \DeclareAcronym{eirp}{
 	short = EIRP,
 	long = effective isotropic radiated power
 }
\DeclareAcronym{gnss}{
	short = GNSS,
	long = global navigation satellite system
}
 \DeclareAcronym{imu}{
 	short = IMU,
 	long = inertial measurement unit
 }

 \DeclareAcronym{ofdm}{
 	short = OFDM,
 	long = orthogonal frequency division multiplexing
 }
 \DeclareAcronym{los}{
 	short = LoS,
 	long = line of sight
 }
 \DeclareAcronym{qam}{
 	short = QAM,
 	long = quadrature amplitude modulation
 }

\DeclareAcronym{pll}{
	short = PLL,
	long = phase-locked loop
}
 \DeclareAcronym{vco}{
 	short = VCO,
 	long = voltage-controlled oscillator
 }
 \DeclareAcronym{lna}{
 	short = LNA,
 	long = low-noise amplifier
 }
 \DeclareAcronym{if}{
 	short = IF,
 	long = intermediate frequency
 }
 \DeclareAcronym{cots}{
 	short = COTS,
 	long = commercial off-the-shelf
 }
\DeclareAcronym{adc}{
	short = ADC,
	long = analog to digital converter
}
\DeclareAcronym{dac}{
	short = DAC,
	long = digital to analog converter
}
\DeclareAcronym{lo}{
	short = LO,
	long = local oscillator
}
 \DeclareAcronym{pcb}{
 	short = PCB,
 	long = printed circuit board
 }
 \DeclareAcronym{mimo}{
 	short = MIMO,
 	long = multiple-input multiple-output
 }
 \DeclareAcronym{simo}{
 	short = SIMO,
 	long = single-input multiple-output
 }
 \DeclareAcronym{mmic}{
 	short = MMIC,
 	long = monolithic microwave integrated circuit
 }
 \DeclareAcronym{daq}{
 	short = DAQ,
 	long = data acquisition
 }
 \DeclareAcronym{ic}{
 	short = IC,
 	long = integrated circuit
 }
 \DeclareAcronym{pa}{
 	short = PA,
 	long = power amplifier
 }

 \DeclareAcronym{ti}{
 	short = TI,
 	long = Texas Instruments
 }
 \DeclareAcronym{adi}{
 	short = ADI,
 	long = Analog Devices
 }

 \DeclareAcronym{roi}{
 	short = ROI,
 	long = region of interest,
 	long-plural-form = regions of interest
 }
 \DeclareAcronym{v2x}{
 	short = V2X,
 	long = vehicle-to-everything
 }
 \DeclareAcronym{av}{
 	short = AV,
 	long = automated vehicle
 }
 \DeclareAcronym{cors}{
 	short = CORS,
 	long = continuously operating reference station
 }
 \DeclareAcronym{mdot}{
 	short = MDOT,
 	long = Michigan Department of Transportation
 }
 \DeclareAcronym{moco}{
 	short = MOCO,
 	long = motion compensation
 }
\DeclareAcronym{sdr}{
	short = SDR,
	long = software-defined radio
}
\DeclareAcronym{fpga}{
	short = FPGA,
	long = field-programmable gate array
}
 \DeclareAcronym{gpio}{
 	short = GPIO,
 	long = general-purpose input/output
 }
 \DeclareAcronym{usrp}{
 	short = USRP,
 	long = Universal Software Radio Peripheral
 }
 \DeclareAcronym{uhd}{
 	short = UHD,
 	long = \ac{usrp} Hardware Driver
 }
 \DeclareAcronym{ntp}{
 	short = NTP,
 	long = network time protocol
 }
 \DeclareAcronym{ptp}{
 	short = PTP,
 	long = precision time protocol
 }
 \DeclareAcronym{lan}{
 	short = LAN,
 	long = local area network
 }
 \DeclareAcronym{wlan}{
 	short = WLAN,
 	long = wireless \ac{lan}
 }
 \DeclareAcronym{lut}{
 	short = LUT,
 	long = lookup table
 }
 \DeclareAcronym{wsn}{
 	short = WSN,
 	long = wireless sensor network
 }
 \DeclareAcronym{mac}{
 	short = MAC,
 	long = media access control
 }
 \DeclareAcronym{pps}{
 	short = PPS,
 	long = pulse-per-second
 }
 \DeclareAcronym{fom}{
 	short = FoM,
 	long = figure of merit
 }
 \DeclareAcronym{uwb}{
 	short = UWB,
 	long = ultra-wideband
 }
 \DeclareAcronym{twtt}{
 	short = TWTT,
 	long = two-way time transfer
 }
 \DeclareAcronym{swap}{
 	short = SWaP,
 	long = {size, weight, and power}
 }
 \DeclareAcronym{cda}{
 	short = CDA,
 	long = {coherent distributed antenna array}
 }
 \DeclareAcronym{ap}{
 	short = AP,
 	long = {access point}
 }

\DeclareAcronym{tof}{
	short = ToF,
	long = {time-of-flight},
 	long-plural-form = {times-of-flight}
}
\DeclareAcronym{gbe}{
	short = GbE,
	long = {Gigabit Ethernet}
}
\DeclareAcronym{gpr}{
	short = GPR,
	long = {ground-penetrating radar}
}
\DeclareAcronym{tdm}{
	short = TDM,
	long = {time-domain multiplexing}
}
\DeclareAcronym{nasa}{
	short = NASA,
	long = National Aeronautics and Space Administration 
}

\ifcsname hlon\endcsname%
  
\else%
  
\fi%

\input{ieee_submission_notice}

\begin{document}

\title{Collaborative Beamforming for Communication Applications Using a Two-Element Fully-Wireless Open-Loop Coherent Distributed Array}

\author{Jason M. Merlo,~\IEEEmembership{Graduate Student Member,~IEEE,}, Jeffrey A. Nanzer,~\IEEEmembership{Senior Member,~IEEE}
\thanks{This effort was sponsored in whole or in part by the Central Intelligence Agency (CIA), through CIA Federal Labs. The U.S. Government is authorized to reproduce and distribute reprints for Governmental purposes notwithstanding any copyright notation thereon. The views and conclusions contained herein are those of the authors and should not be interpreted as necessarily representing the official policies or endorsements, either expressed or implied, of the Central Intelligence Agency.}
\thanks{J. M. Merlo and J. A. Nanzer are with the Department of Electrical and Computer Engineering, Michigan State University, East Lansing, MI 48824 USA (e-mail: merlojas@msu.edu; nanzer@msu.edu).}}

\maketitle
%
%%%%%%%%%%%%%%%%%%%%%%%%%%%%%%%%%%%%%%%%%%%%%%%%%%%%%%%%%%%%%%%%%%%%%%%%%%%%%

\begin{abstract}
In this work we demonstrate a proof of concept of a fully-wireless two-node open-loop coherent distributed communication system and evaluate its performance by transmitting QPSK , 64--, and 256--\acs{qam} constellations at a symbol rate of 2\;MBd over a 58\;m link in an urban environment. The system is implemented in a distributed manner with on-node processing using \acfp{sdr} and wireless internode communication to share coordination information and does not rely on external time or frequency references such as the \acf{gnss}. In each experiment $\mathbf{\sim}$100 messages were transmitted and a mean coherent gain of 0.936 was achieved across all measurements with a mean symbol error ratio of below $\mathbf{1.4\times10^{-4}}$  achieved up to 64-\ac{qam}, demonstrating a reliable bandwidth of up to 12\;Mbps.
\end{abstract}
\acresetall

\section{Introduction}

Interest in wirelessly coordinated open-loop \acp{cda} has been increasing in recent years due to their promise to enhance existing communication and sensing systems by providing increased scalability, increased reconfigurability, and increased robustness to element failure, with respect to traditional monolithic antenna arrays, since elements can be freely added, removed, and relocated at runtime to adapt to changing operational requirements~\cite{nanzer2021distributed-arrays}. Applications for \acp{cda} include distributed sensing~\cite{prager2020wireless,merlo2024distributed-radar,werbunat2024repeater-for-coherent-radar}, microwave imaging~\cite{luzano2024distributed-microwave-interferometer,nusrat2024distributed-repeaters}, wireless power transfer~\cite{brunet2024wireless-power-transfer,choi2018distributed-wireless-power-transfer}, and terrestrial and space-based communication~\cite{abari2015airshare,holtom2024discobeam,quadrelli2019distributed-swarm}\cite[\S TX05.2.6]{nasa2020taxonomy}. 

While the concept of \acp{cda} has been around for decades~\cite{ochiai2005collaborative-beamforming}, many experimental demonstrations have employed closed-loop beamforming using feedback from a designated receiver node~\cite{bidgare2012implementation, quitin2013distributed-transmit-beamforming} which limits the application to cooperative communication scenarios; however, recent advances in wireless system coordination at the carrier wavelength~\cite{prager2020wireless,mghabghab2021frequency-syntonization,merlo2022wireless,kenney2024distributed-frequency-and-phase-synchronization,aguilar2024uncoupled-digital-radars} have enabled experimental measurements of arrays using open-loop topologies where the destination does not provide feedback to the transmit array. While there have been several notable experimental demonstrations of open-loop \acp{cda} investigating their use in sensing and communication applications~\cite{prager2020wireless,merlo2024distributed-radar,holtom2024discobeam,werbunat2024repeater-for-coherent-radar}, the majority of experimental demonstrations to this point have mainly focused sending \ac{cw} tones to assess the array gain, but have not implemented applications running atop the array to identify and address challenges unique to \acp{cda}, in practice.

\begin{figure}
	\centering
	\includegraphics[width=0.35\columnwidth]{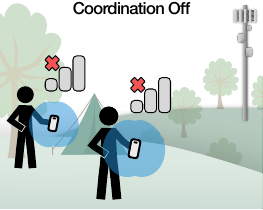}\hspace{0.1\columnwidth}
	\includegraphics[width=0.35\columnwidth]{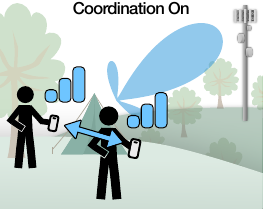}\\
	(a)\hspace{0.325\columnwidth}\hspace{0.1\columnwidth}(b)
	\caption{Collaborative beamforming used in a communication scenario. In (a), an individual \acs{ue} cannot access the \acf{bs} due to a low signal gain. In (b), the \acsp{ue} coordinate wirelessly to perform collaborative beamforming to increase signal gain and establish communication with the \acs{bs}.}
	\label{ov1}
\end{figure}

The goal of this work is to demonstrate a two-node distributed communication system transmitter. A motivating example for this work is provided in Fig. \ref{ov1}, illustrating a \ac{cda} for use in a mobile communication application where  two or more user devices (\acsp{ue}) coordinate wirelessly to extend their range via collaborative beamforming. We build on the time and frequency coordination techniques described in~\cite{mghabghab2021frequency-syntonization} and~\cite{merlo2022wireless} in a system similar to those described in~\cite{merlo2022wireless,merlo2024distributed-radar}, but with performance improve ments in two key areas: the system no longer requires a \acf{gnss} receiver for coarse time alignment; and the synchronization epoch duration has been reduced from $\sim$\SI{100}{\milli\second} to $\sim$\SI{10}{\micro\second}, greatly improving the resiliency to dynamic environmental multipath.  In these experiments, we use \ac{qam} constellations, first with a low order \ac{qpsk}, then with 64-- and 256--\ac{qam} constellations.  The added gain of the array from the second node is evaluated using a series of \ac{cw} pulses in the preamble, the carrier phase alignment is evaluated through orthogonal \ac{zc} sequences transmitted in the preamble, and the \ac{evm} and \ac{ser} are evaluated based on the transmission of pseudo-random sequences in the \ac{qam} payload waveforms.

%%%%%%%%%%%%%%%%%%%%%%%%%%%%%%%%%%%%%%%%%%%%%%%%%%%%%%%%%%%%%%%%%%%%%%%%%%%%%

\section{Distributed Transmit Beamforming Communication System}

\begin{figure}
	\centering
	\includegraphics[width=0.8\columnwidth]{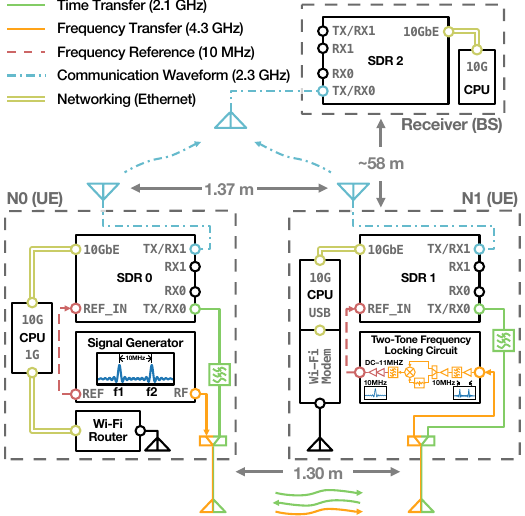}
	\caption{Distributed transmit system schematic.  The system used two transmitting nodes which were wirelessly coordinated and time and frequency to beamform communication waveforms.  %Each node used an Ettus Research X310 \ac{sdr} and contained its own computer for locally processing the sampled data. Wireless frequency locking was accomplished using a two-tone wireless frequency transfer technique, and time synchronization was accomplished digitally using a high-accuracy two-way time transfer technique; nodes communicated beamforming information and exchanged \ac{twtt} timestamps using Wi-Fi. 
	An unsynchronized receive node was located 58\;m down-range and recorded data for demodulation.}
	\label{schematic}
\end{figure}

The primary challenge with wirelessly coordinated distributed systems is ensuring that they sum coherently at the receiver. To accomplish this, the time, frequency, and phase offsets between each system must be estimated and compensated for down to a fraction of a carrier wavelength~\cite{nanzer2021distributed-arrays}.  To model this, we represent the transformation between the global true time $t$ and the local time at any node $n$ as
\begin{equation}
\label{local-time}
    T^{(n)}(t) = \alpha^{(n)}(t)t + \delta^{(n)}(t) + \nu^{(n)}(t)
\end{equation}
where $\alpha^{(n)}(t)$ is the time-varying relative frequency scaling coefficient, $\delta^{(n)}(t)$ is the time-varying time bias, and $\nu^{(n)}(t)$ is a zero-mean noise term due to device properties such as thermal noise, flicker noise, shot noise, etc.  It should be noted that, in general the transmit and receive pathways of a radio will have separate functions $T$ to represent their offset due to having unique time and phase delays in their respective \ac{rf} pathways.  The \ac{rf} carrier can be modeled by
\begin{equation}
    \Phi_\mathrm{RF}^{(n)}(t)=\exp{\left\{j\left[2.0\pi\, f_\mathrm{RF}\,T^{(n)}(t)+\phi_{0}^{(n)}\right]\right\}}
\end{equation}
where $f_\mathrm{RF}$ is the intended carrier frequency and $\phi_0$ is the initial phase of the \ac{lo}.

To ensure the transmitted waveforms sum coherently at the receiver, the transmitters need to estimate and compensate for: relative frequency offset between systems $\alpha^{(n)}$, relative time offset between systems $\delta^{(n)}$, and the inter-arrival phases of the waveforms transmitted from each node.  Assuming the systems are syntonized---aligned in frequency---the inter-arrival phases are a function of the initial \ac{lo} phase $\phi_0^{(n)}$, the phase delay induced by $\delta^{(n)}$, and the relative locations of the transmitting and receiving nodes. In this work we establish the initial phase offset of each \ac{rf} front-end and relative node locations by performing an initial far-field calibration to the receiving node.  Thus, only $\delta^{(n)}$ must be continuously estimated online.

To align the \ac{lo} frequencies between systems and compensate for $\alpha^{(n)}$, a two-tone \ac{cw} reference tone was transmitted from \ac{ue} node 0 (the primary node) and demodulated using self-mixing circuit described in~\cite{abari2015airshare,mghabghab2021frequency-syntonization}; this provides a fixed relative phase between nodes in static environments. To align the time and phase, we use the technique described in~\cite{merlo2022wireless}, which uses a \ac{twtt} process with a \ac{qls} refinement process and \ac{lut}-based bias compensation phase. To compensate for the time and calibrated phase, a delay and phase is applied to the sampled baseband waveform. The baseband compensation waveform to be added at each node is
\begin{align}
\begin{split}
	&s_\mathrm{comp}^{(n)}(t) = \exp\left\{-j\left[2.0\,\pi\,f_\mathrm{RF}\tilde{\delta}^{(n)}(t)+\tilde{\phi}_0\right]\right\}	
\end{split}
\end{align}
where $\tilde{(\cdot)}$ represents an estimated quantity. Thus, the total waveform transmitted from each node after applying compensation and baseband modulation can be modeled as
\begin{align}
\begin{split}
	&s^{(n)}(t) = \Phi_\mathrm{RF}^{(n)}(t)\,	s_\mathrm{comp}^{(n)}(t) \,s_\mathrm{m}\!\left(T^{(n)}(t)\right)
\end{split}
\end{align}
where $s_\mathrm{m}(t)$ is the desired baseband message to be transmitted---in this case a \ac{qam} message.

%%%%%%%%%%%%%%%%%%%%%%%%%%%%%%%%%%%%%%%%%%%%%%%%%%%%%%%%%%%%%%%%%%%%%%%%%%%%%

\section{Experimental Configuration and Results}

\begin{figure}
	\centering
	\includegraphics[width=0.8\columnwidth]{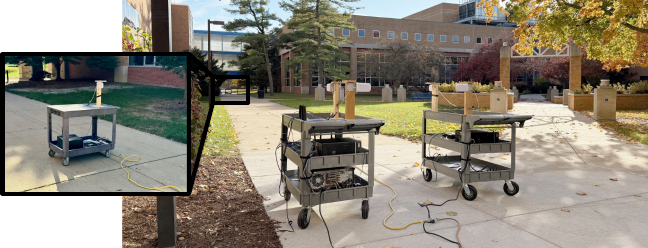}
	\caption{Photograph of the distributed transmit array (foreground) and receive node (inset).  Node 0 is located on the left, and Node 1 is located on the right. Log-periodic antennas on wooden masts were used for time-frequency coordination as well as beamforming and can be seen on top of the carts.}
	\label{setup}
\end{figure}

The experimental system is shown schematically in Fig. \ref{schematic} and photographed in Fig. \ref{setup}. Each \ac{ue} node was separated by $\sim$\SI{1.3}{\meter} and contained an Ettus Research X310 \ac{sdr} connected to a Dell Optiplex 7080 desktop. %with an Intel i7-10700 and 16 GB of DDR4 memory.% using Ubuntu 22.04 with GNU Radio 3.10, UHD 4.7, and Python 3.10.  
Each node communicated over TCP/IP using ZeroMQ via Wi-Fi.% using channel 165 (\SI{5825}{\mega\hertz}) to share synchronization timestamps and beamforming information.

The time and frequency coordination waveforms were transmitted between nodes using shared log-periodic antennas. The primary node N0 contained a signal generator which acts as the primary \ac{lo} for the system; the \SI{10}{\mega\hertz} output was used to directly discipline the \ac{lo} on \ac{sdr} 0 while a two-tone \ac{cw} waveform was transmitted at \SI{4305}{\mega\hertz} as reference for \ac{ue} N1.  N1 demodulated the frequency reference using the two-tone frequency locking circuit and use the \SI{10}{\mega\hertz} output to discipline the \ac{lo} on \ac{sdr} 1. The time transfer operation was performed using channel 0 on each \ac{sdr} at \SI{2100}{\mega\hertz}.  The initial fine time--phase acquisition process utilized TCP/IP packets transmitted between nodes to query the local times on each radio which provided an initial estimate on the order of \SI{10}{\milli\second}; after initial network-based time alignment, the nodes would send a pulsed two-tone waveform synchronization waveform with a low tone separation (\SI{1}{\mega\hertz}) at a low sample rate (\SI{10}{\mega Sa/\second}) with a wide \ac{tdm} window duration, then iteratively increase the sample rate and tone separation while reducing the \ac{tdm} window to values of \SI{20}{\mega\hertz}, \SI{200}{\mega Sa/\second}, and \SI{10}{\micro\second}; in all cases, pulse durations of \SI{2}{\micro\second} were used.

\begin{figure}
	\centering
	\includegraphics{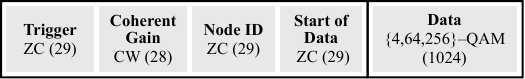}
	\caption{Message structure.  Preamble contained four sequences, three of which were 29-symbol \acf{zc} sequences, and the fourth was a set of \acf{cw} pulses intended to evaluate carrier phase coherence of the system. Three data modulation schemes were evaluated: \ac{qpsk}, 64--\ac{qam}, and 256--\ac{qam}.}
	\label{message}
\end{figure}

The beamforming waveform was transmitted from each node using an 8-dBi log-periodic antenna at \SI{2300}{\mega\hertz} at a sample rate of \SI{200}{\mega Sa/\second} with a symbol rate of \SI{2}{\mega Bd}. A \ac{rrc} window was applied to each symbol with a duration of 41 symbols and a roll-off factor of 0.3. The message used a four-part preamble consisting of a trigger, coherent gain estimation pattern, node identifier, and start of data indicator, shown in Fig. \ref{message}.  The trigger, node identifier, and start of data indicator were all 29-symbol \ac{zc} sequences where the node identifiers were unique for each \ac{ue}. The coherent gain estimation waveform consisted of 4-symbol long constant-phase pulses where first \ac{ue} N0 would transmit, then \ac{ue} N1 would transmit, then both \acp{ue} would transmit simultaneously so that the total coherent gain $G_\mathrm{c}$ could be estimated, where $G_\mathrm{c}$ is the fraction of the measured combined pulse's amplitude relative to the ideal summation of each of the individual pulse amplitudes. The node identifier was used for calibration and performance evaluation since the individual nodes magnitude and phase could be estimated by matching filtering to the orthogonal \ac{zc} sequence. The payload of the message contained 1024 symbols of pseudo-random \ac{qam} data using \ac{qpsk}, 64--, and 256--\ac{qam} schemes.  
%\begin{figure}
%	\centering
%	\includegraphics{figures/qam_processing_single_column}
%	\caption{Receiver message demodulation process.}
%	\label{demodulator}
%\end{figure}
The receiver \acf{bs} node was placed \SI{58}{\meter} downrange and used a similar desktop computer and \ac{sdr} configuration as the transmitters, but because it was not time synchronized with the transmitter \acp{ue}, it was run in a continuous streaming mode which limited the receive sample rate to \SI{50}{\mega Sa/\second}.  A matched filter for the trigger preamble was run on the data in real-time to save the waveforms for offline evaluation and demodulation.  %The demodulation process used is illustrated in Fig. \ref{demodulator}.

\begin{figure}[tb]
	\centering
	\includegraphics[width=0.9\columnwidth]{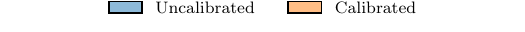}
	\includegraphics[width=0.9\columnwidth]{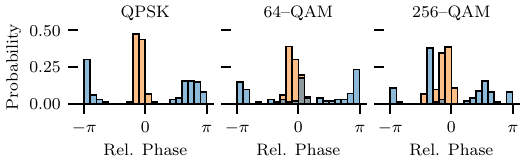}
	\includegraphics[width=0.9\columnwidth]{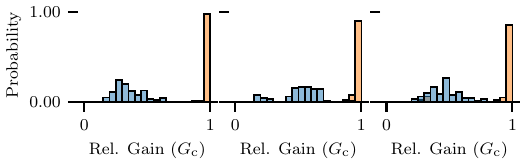}
	\includegraphics[width=0.9\columnwidth]{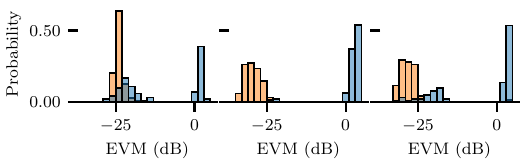}
	\includegraphics[width=0.9\columnwidth]{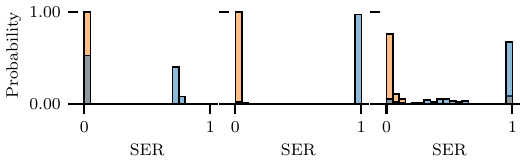}
	\caption{Measured performance metrics for \ac{qpsk}, 64--, and 256--\ac{qam}. Top to bottom are relative inter-arrival phase, coherent gain relative to ideal summation, \acf{evm}, \acf{ser}.}
	\label{results}
\end{figure}

The data collection process for each constellation type consisted of both an uncalibrated control measurement where front-end calibrations were not applied (but time and frequency coordination was still operational) and a calibrated measurement where the front-end was calibrated ahead of time to the receiver location using the node identifier preamble to estimate inter-node time and phase of arrival corrections; after the calibration process the program was restarted and the calibrated measurements were taken.  For each modulation order 100--110 messages were collected and used to compute the inter-node carrier phase offsets, coherent gain, \ac{evm}, and \ac{ser} at the receiver.  A summary of these statistics is shown in Fig. \ref{results}.
The relative inter-arrival phase is shown in the first row and illustrates the impact of proper phase alignment between the nodes; in the calibrated case, the phase due to static differences in the \ac{rf} front-end, transmission lines, antenna patterns, and propagation distances are corrected. The second row illustrates the improvement in coherent gain performance after calibration which resulted in a mean coherent gain improvement from 0.442 for the uncalibrated case to 0.963 after calibration.  The third and fourth rows shown the \ac{evm} and \ac{ser}, respectively, and illustrate the data demodulation performance improvement after calibration.  In both the \ac{qpsk} and 64--\ac{qam} cases, the \ac{ser} was below $1.4\times10^{-4}$, while the \ac{ser} of the 256--\ac{qam} modulation was higher, possibly due to slight signal compression, and several unrecoverable messages due to \ac{cfo} estimation failure which was not observed in other modulation orders, which may have been caused by out of band interference.

%\begin{figure}
%	\centering
%	\includegraphics[width=\columnwidth]{../analysis/results/241028_221306_QAM256_calibrated/constellation_file6_90_195.89392854.pdf}
%%	\includegraphics[width=\columnwidth]{../analysis/results/241028_221306_QAM256_calibrated/eye_file6_90_195.89392854.pdf}
%%	\includegraphics[width=\columnwidth]{../analysis/results/241028_221835_QAM256_uncalibrated/constellation_file6_37_147.06678524.pdf}
%	\caption{Measured 256--\ac{qam} constellation diagram over one message frame after calibration.}
%\end{figure}

%%%%%%%%%%%%%%%%%%%%%%%%%%%%%%%%%%%%%%%%%%%%%%%%%%%%%%%%%%%%%%%%%%%%%%%%%%%%%

\section{Conclusion}

In this work, we demonstrate initial measurements of a \ac{cda} transmitter for collaborative communication beamforming applications in an urban environment.  We show excellent performance for \ac{qpsk} and 64--\ac{qam} constellations and promising results for higher order modulation schemes with further tuning.  A coherent gain of 0.936 was achieved across all measured waveforms, and a \ac{ser} of below $1.4\times10^{-4}$ was obtained up to 64--\ac{qam} enabling raw bitrates of up to \SI{12}{\mega bp\second}. These results show encouraging performance and motivate further work studying scaling array sizes and symbol rates for future efforts.

%%%%%%%%%%%%%%%%%%%%%%%%%%%%%%%%%%%%%%%%%%%%%%%%%%%%%%%%%%%%%%%%%%%%%%%%%%%%%
%\newpage
  
\bibliographystyle{IEEEtran}

\bibliography{reference_library}

% Generated by IEEEtran.bst, version: 1.14 (2015/08/26)
\begin{thebibliography}{10}
\providecommand{\url}[1]{#1}
\csname url@samestyle\endcsname
\providecommand{\newblock}{\relax}
\providecommand{\bibinfo}[2]{#2}
\providecommand{\BIBentrySTDinterwordspacing}{\spaceskip=0pt\relax}
\providecommand{\BIBentryALTinterwordstretchfactor}{4}
\providecommand{\BIBentryALTinterwordspacing}{\spaceskip=\fontdimen2\font plus
\BIBentryALTinterwordstretchfactor\fontdimen3\font minus \fontdimen4\font\relax}
\providecommand{\BIBforeignlanguage}[2]{{%
\expandafter\ifx\csname l@#1\endcsname\relax
\typeout{** WARNING: IEEEtran.bst: No hyphenation pattern has been}%
\typeout{** loaded for the language `#1'. Using the pattern for}%
\typeout{** the default language instead.}%
\else
\language=\csname l@#1\endcsname
\fi
#2}}
\providecommand{\BIBdecl}{\relax}
\BIBdecl

\bibitem{nanzer2021distributed-arrays}
J.~A. Nanzer, S.~R. Mghabghab, S.~M. Ellison, and A.~Schlegel, ``Distributed phased arrays: Challenges and recent advances,'' \emph{IEEE Transactions on Microwave Theory and Techniques}, vol.~69, no.~11, pp. 4893--4907, 2021.

\bibitem{prager2020wireless}
S.~Prager, M.~S. Haynes, and M.~Moghaddam, ``Wireless subnanosecond rf synchronization for distributed ultrawideband software-defined radar networks,'' \emph{IEEE Transactions on Microwave Theory and Techniques}, vol.~68, no.~11, pp. 4787--4804, 2020.

\bibitem{merlo2024distributed-radar}
J.~M. Merlo, S.~Wagner, J.~Lancaster, and J.~A. Nanzer, ``Fully wireless coherent distributed phased array system for networked radar applications,'' \emph{IEEE Microwave and Wireless Technology Letters}, vol.~34, no.~6, pp. 837--840, 2024.

\bibitem{werbunat2024repeater-for-coherent-radar}
D.~Werbunat, B.~Woischneck, J.~Lerch, B.~Schweizer, R.~Michev, C.~Bonfert, J.~Hasch, and C.~Waldschmidt, ``Multichannel repeater for coherent radar networks enabling high-resolution radar imaging,'' \emph{IEEE Transactions on Microwave Theory and Techniques}, vol.~72, no.~5, pp. 3247--3259, 2024.

\bibitem{luzano2024distributed-microwave-interferometer}
D.~Luzano, J.~M. Merlo, D.~Chen, J.~R. Colon-Berrios, A.~Bhattacharyya, and J.~A. Nanzer, ``A distributed microwave correlation interferometer for fourier domain imaging using wireless time and frequency coordination,'' in \emph{2024 IEEE International Symposium on Antennas and Propagation and INC/USNC-URSI Radio Science Meeting}, 2024, pp. 1455--1456.

\bibitem{nusrat2024distributed-repeaters}
T.~Nusrat and S.~Vakalis, ``Addressing specularity: Millimeter-wave radar with distributed repeater apertures,'' \emph{IEEE Transactions on Microwave Theory and Techniques}, pp. 1--10, 2024.

\bibitem{brunet2024wireless-power-transfer}
J.~Brunet, A.~Ayling, and A.~Hajimiri, ``Transmitarrays for wireless power transfer on earth and in space,'' \emph{IEEE Journal of Microwaves}, pp. 1--12, 2024.

\bibitem{choi2018distributed-wireless-power-transfer}
K.~W. Choi, A.~A. Aziz, D.~Setiawan, N.~M. Tran, L.~Ginting, and D.~I. Kim, ``Distributed wireless power transfer system for internet of things devices,'' \emph{IEEE Internet of Things Journal}, vol.~5, no.~4, pp. 2657--2671, 2018.

\bibitem{abari2015airshare}
O.~Abari, H.~Rahul, D.~Katabi, and M.~Pant, ``Airshare: Distributed coherent transmission made seamless,'' in \emph{2015 IEEE Conference on Computer Communications (INFOCOM)}, 2015, pp. 1742--1750.

\bibitem{holtom2024discobeam}
J.~Holtom, O.~Ma, A.~Herschfelt, I.~Lenz, Y.~Li, and D.~W. Bliss, ``Distributed coherent mesh beamforming (discobeam) for robust wireless communications,'' \emph{IEEE Transactions on Wireless Communications}, vol.~23, no.~11, pp. 15\,814--15\,828, 2024.

\bibitem{quadrelli2019distributed-swarm}
M.~B. Quadrelli, R.~Hodges, V.~Vilnrotter, S.~Bandyopadhyay, F.~Tassi, and S.~Bevilacqua, ``Distributed swarm antenna arrays for deep space applications,'' in \emph{2019 IEEE Aerospace Conference}, 2019, pp. 1--15.

\bibitem{nasa2020taxonomy}
``2020 {NASA} technology taxonomy,'' National Aeronautics and Space Administration, Tech. Rep. HQ-E-DAA-TN76545, Jan. 2020.

\bibitem{ochiai2005collaborative-beamforming}
H.~Ochiai, P.~Mitran, H.~Poor, and V.~Tarokh, ``Collaborative beamforming for distributed wireless ad hoc sensor networks,'' \emph{IEEE Transactions on Signal Processing}, vol.~53, no.~11, pp. 4110--4124, 2005.

\bibitem{bidgare2012implementation}
P.~Bidigare, M.~Oyarzyn, D.~Raeman, D.~Chang, D.~Cousins, R.~O'Donnell, C.~Obranovich, and D.~R. Brown, ``Implementation and demonstration of receiver-coordinated distributed transmit beamforming across an ad-hoc radio network,'' in \emph{2012 Conference Record of the Forty Sixth Asilomar Conference on Signals, Systems and Computers (ASILOMAR)}, 2012, pp. 222--226.

\bibitem{quitin2013distributed-transmit-beamforming}
F.~Quitin, M.~M.~U. Rahman, R.~Mudumbai, and U.~Madhow, ``A scalable architecture for distributed transmit beamforming with commodity radios: Design and proof of concept,'' \emph{IEEE Transactions on Wireless Communications}, vol.~12, no.~3, pp. 1418--1428, 2013.

\bibitem{mghabghab2021frequency-syntonization}
S.~R. Mghabghab and J.~A. Nanzer, ``Open-loop distributed beamforming using wireless frequency synchronization,'' \emph{IEEE Transactions on Microwave Theory and Techniques}, vol.~69, no.~1, pp. 896--905, 2021.

\bibitem{merlo2022wireless}
J.~M. {Merlo}, S.~R. {Mghabghab}, and J.~A. {Nanzer}, ``Wireless picosecond time synchronization for distributed antenna arrays,'' \emph{IEEE Transactions on Microwave Theory and Techniques}, vol.~71, no.~4, pp. 1720--1731, Dec. 2022.

\bibitem{kenney2024distributed-frequency-and-phase-synchronization}
R.~H. Kenney, J.~G. Metcalf, and J.~W. McDaniel, ``Wireless distributed frequency and phase synchronization for mobile platforms in cooperative digital radar networks,'' \emph{IEEE Transactions on Radar Systems}, vol.~2, pp. 268--287, 2024.

\bibitem{aguilar2024uncoupled-digital-radars}
J.~Aguilar, D.~Werbunat, V.~Janoudi, C.~Bonfert, and C.~Waldschmidt, ``Uncoupled digital radars creating a coherent sensor network,'' \emph{IEEE Journal of Microwaves}, vol.~4, no.~3, pp. 459--472, 2024.

\end{thebibliography}

\end{document}